# Advanced Data Processing in the Business Network System

Daniel Ritter

*Abstract*—**The discovery, representation and reconstruction of Business Networks (BN) from Network Mining (NM) raw data is a difficult problem for enterprises. This is due to huge amounts of e.g. complex business processes within and across enterprise boundaries, heterogeneous technology stacks, and fragmented data. To remain competitive, visibility into the enterprise and partner networks on different, interrelated abstraction levels is desirable.**

**We show the query and data processing capabilities of a novel data discovery, mining and network inference system, called Business Network System (BNS) that reconstructs the BN - integration and business process networks - from raw data, hidden in the enterprises' landscapes. The paper covers both the foundation and the key data processing characteristics features of BNS, including its underlying technologies, its overall system architecture, and data provenance approach.**

## I. Introduction

Enterprises are part of value chains consisting of business processes connecting intra- and inter-enterprise participants. The network that connects these participants with their technical, social and business relations is called a *Business Network* (BN). Even though this network is very important for the enterprise, there are few - if any - people in the organization who understand this network as the relevant data is hidden in heterogeneous enterprise system landscapes. Yet simple questions about the network (e.g., which business processes require which interfaces, which integration artifacts are obsolete) remain difficult to answer, which makes the operation and lifecycle management like data migration, landscape optimization and evolution hard and more expensive increasing with the number of the systems. To change that, *Network Mining* (NM) systems are used to discover and extract raw data [13] - be it technical data (e.g., configurations of integration products like *Enterprise Service Bus* (ESB) [8]) or business data (e.g., information about a supplier in a *Supplier Relationship Management* (SRM) product). The task at hand is to provide a system, that automatically discovers and reconstructs the "as-is" BN from the incomplete, fragmented, cross-domain NM data and make it accessible for visualization and analysis.

Previous work on NM systems [13], their extension towards a holistic management of BN [15] and a cloud-based reference architecture [16] provide a comprehensive, theoretical and practical foundation on how to build a system suited to this task, called the *Business Network System* (BNS).

In this work we discuss the data processing and provenance requirements of the BNS and shed light into its internal mechanics. The major contributions of this work are (1) a sound list of the most important requirements of the data processing in the BNS, building on previous work, (2) a data provenance approach suitable for these requirements, and (3) a system implementing this architecture for continuous and scalable end-to-end network query, traversal and update processing based on the data transformation and provenance approach. We applied our system to several real-world enterprise landscapes.

Section II guides from the theoretical work conducted in the area of NM [13], *Business Network Management* (BNM) [15], and its reference architecture [16] to the real-world query and data processing and provenance requirements and capabilities of a BNS (refers to (1)). Section III provides an overview of BNS's data transformations, provenance and data processing, including query and update processing (refers to (2)) and sketches a high-level view on the system's architecture (refers to (3)). Section IV reviews and discusses related work and systems that influenced BNS. Section V concludes the paper and lists some of the future work.

## II. The Business Network System

The BN consists of a set of interrelated perspectives of domain networks (e.g., business process, integration, social), that provide a contextualized view on which business processes (i.e., business perspective) are currently running, implemented on which integration capabilities (i.e., integration perspective) and operated by whom (i.e., social perspective). To compute the BN, Network Mining (NM) systems automatically discover raw data from the enterprise landscapes [13]. These conceptual foundations are extended to theoretically ground the new BN data management domain [15].

The fundamental requirements and capabilities of a BNS are derived from the theoretical foundations in previous as well as related work are extensively discussed in [16]. In a nutshell they cover (a) the (semi-)automatic discovery of data within the enterprise landscapes and cloud applications, (b) a common, domain independent, network inference model, (c) the transformation of the domain data into this model, (d) a scalable, continuously running and declaratively

Daniel Ritter is with the Technology Development at the SAP AG, Walldorf, BW 69190, Germany (e-mail: daniel.ritter@sap.com).

programmable inference system, (e) the cross-domain and enterprise/ tenant reconstruction, (f) the ability to check the data quality and compliance to the inference model, and (g) the visualization of different perspectives (i.e., views) on the BN (e.g., business process, integration). When starting with a system, which fulfills these requirements ((a)-(g)), the query and data processing aspects of the BNS summarize to the following:

1. *REQ-1* The client API shall allow remote access as well as scalable query, traversal and full-text search across the interconnected BN perspectives (e.g., through index creation) (from [14]).
2. *REQ-2* The (remote) API shall provide a standard exchange and visualization format (i.e., common standard in NM related areas like BPM) (from [12] and [14]).
3. *REQ-3* The user shall be able to enrich (e.g., labeling, grouping) and to enhance the BN data (e.g., adding/ removing nodes and edges) through the client API, while delta-changes from existing data sources and initial loads from new data sources are merged into the existing BN (from [15]).
4. *REQ-4* Through the whole system, the data origin shall be tracked through all transformations from the source to the BN (i.e., data provenance). This shall allow for continuous data source integration, user enrichments/ enhancements as well as possible re-deployment from the BN to the data sources (from [15]).
5. *REQ-5* The source data shall be available at all times for continuous re-computation of the network (i.e., even if the original source is not accessible for a while) (from [15] and *REQ-4*).

To sketch an idea on what these requirements mean for the query and data processing of a BNS, Figure 1 helps to provide a high-level map to locate the core data processing capabilities of our BNS. On the bottom is reality - a mix of business process, social and integration artifacts stored in databases, packaged applications, system landscape directories (e.g., SAP SLD [18]), middleware systems (e.g., SAP PI [17]), documents/ files, application back-ends, and so on. When pointed to an enterprise data source through configuration by a domain expert, the BNS introspects the source's metadata (e.g., WSDL file for Web service), discovers and transforms the domain data to a common representation (see (a)). Other functional and queryable data sources are similarly processed.

The center of Figure 1 shows the core elements of a NM system, theoretically discussed in [13] and [15], which computes the perspectives of the BN for client access. After the loaded data has been checked for conformance to the inference model it is stored as raw data for the continuously running network reconstruction programs using logic programming (i.e., our approach uses Datalog due to the rationale in [16]). Since BN reconstruction works on cross-domain and the enterprise data, and (cloud) applications want to access the BN data, the NM-part of the system is located in the public cloud, while the discovery-part is located in the enterprise system landscapes. That means, the data sources are highly distributed and the permanent, efficient access is not guaranteed. For that, the source data is copied to the Business Network Server by a set of protocol adapters. The data is stored as raw data, but linked to its original source for later reference (see *REQ-5*).

The computation of the network results in interrelated network perspectives, which are accessed by the clients for network visualization, simulation or analytics (see (g), *REQ-1* and *REQ-2*). User updates are brought into the system through the client API and are visible (at least) for the authors (see *REQ-3*). In each of the steps the data provenance is updated to preserve the path to the origin of the data from the client queries to the source models (see *REQ-4*). Together with *REQ-5*, an end-to-end lineage from the original source data artifacts to the visualized instances of the computed network shall be possible.

## III. Data Processing In the Business Network System

### A. *Query Processing Overview*

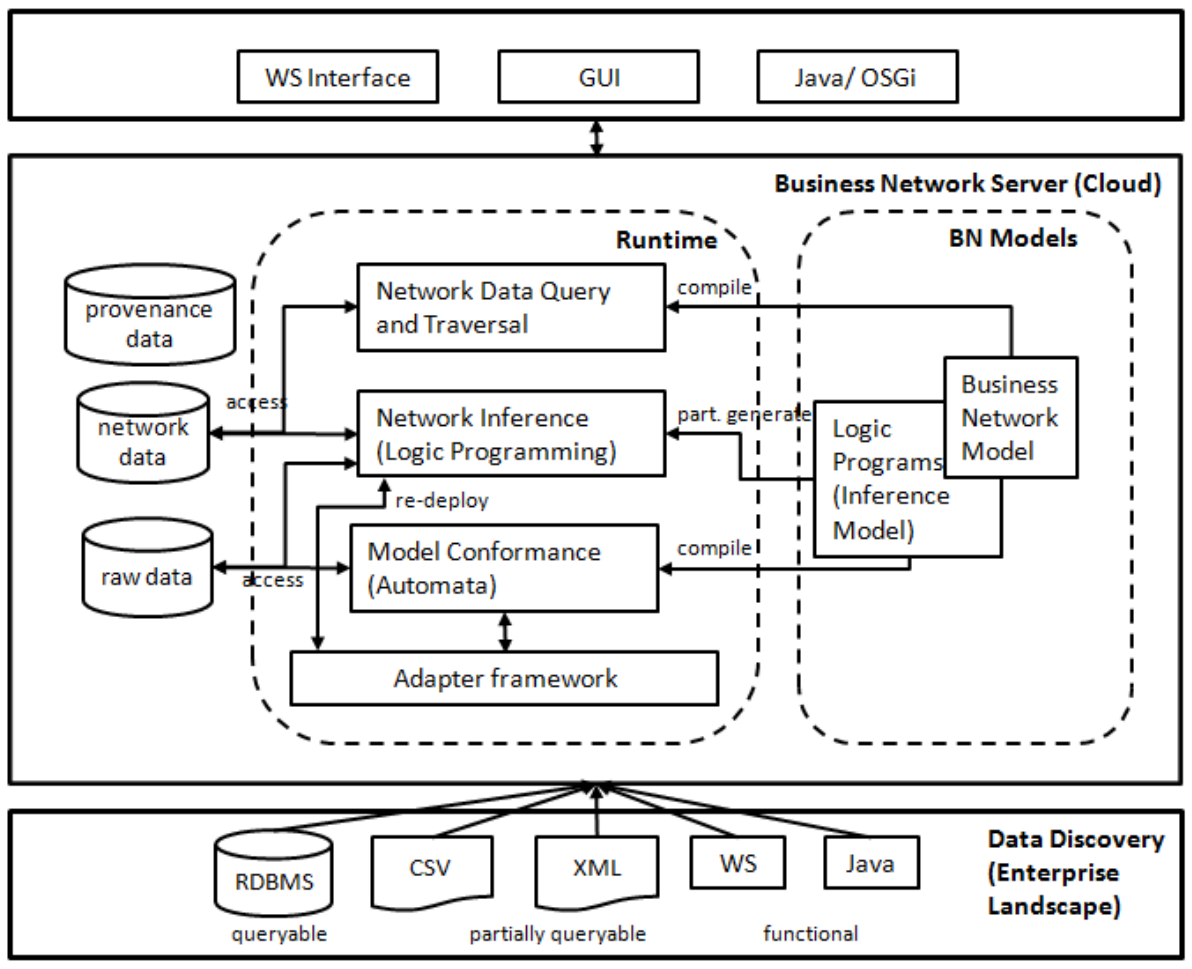

Fig. 1. Overview of the Business Network architecture

Figure 1 provides an overview of BNS's internal architecture. At the bottom of the figure are various types of data sources from which BNS can receive data (the default is the XML upload). During the upload of the data it is checked for conformance with the inference model (shown in the center) by an automata-based runtime, compiled and configured from the model. When the check is successful, the data is stored in the knowledge base as raw data (see *REQ-5*). The network inference programs (partially generated from the model) run decoupled from the upload or inbound processing, while working on snapshots of the raw data. The inference result is stored as network data, which automatically updates the indices for full-text search, query and traversal on the data. The client requests are again handled independent of the inbound processing and network inference only on the current BN. The access layer is based on a flexible resource representation from [14], which adapts to changes in the BN model automatically (see *REQ-1*, *REQ-2*). The BNS provides its own network visualization for the different perspectives and contextualization. Figure 2 shows an excerpt of a real-world integration network perspective and the drill-down to the message flow details (see Figure 3).

In addition, a Java/ OSGi declarative service interface and a HTTP/JSON remote API are provided to build own UIs, enrich or enhance the computed "as-is" network, and build applications for network analytics, optimization or monitoring (see *REQ-1*).

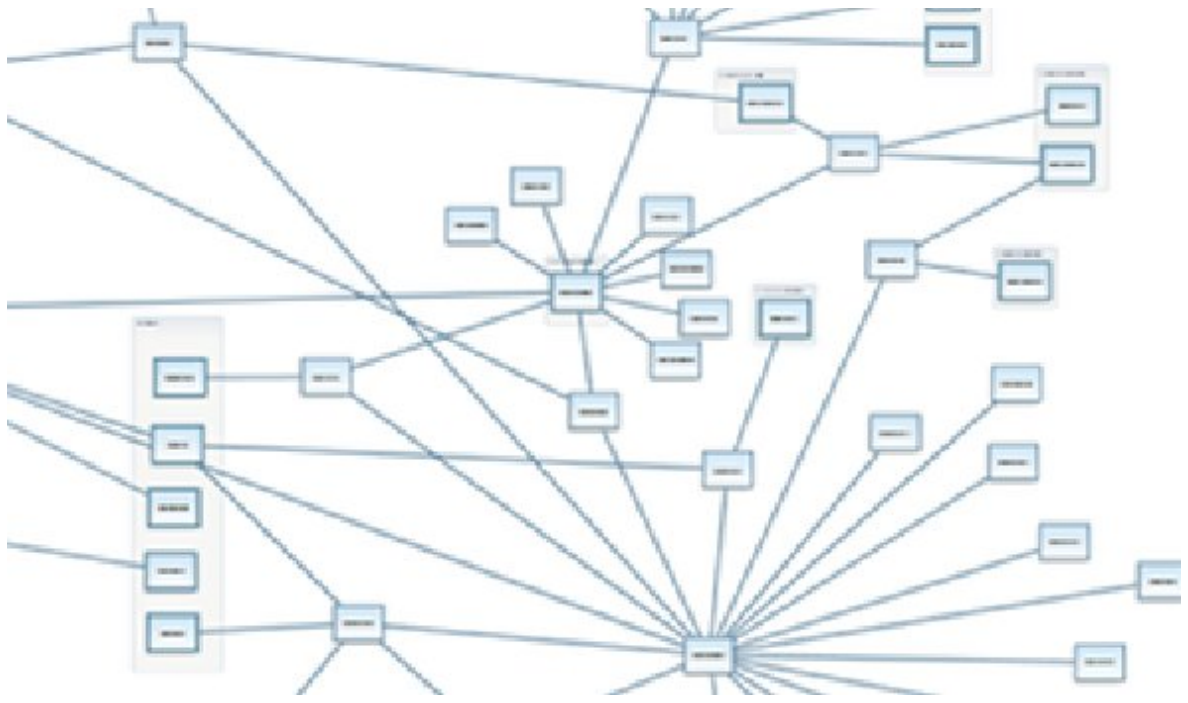

Fig. 2. Integration network visualization showing a high-level view on an integration network

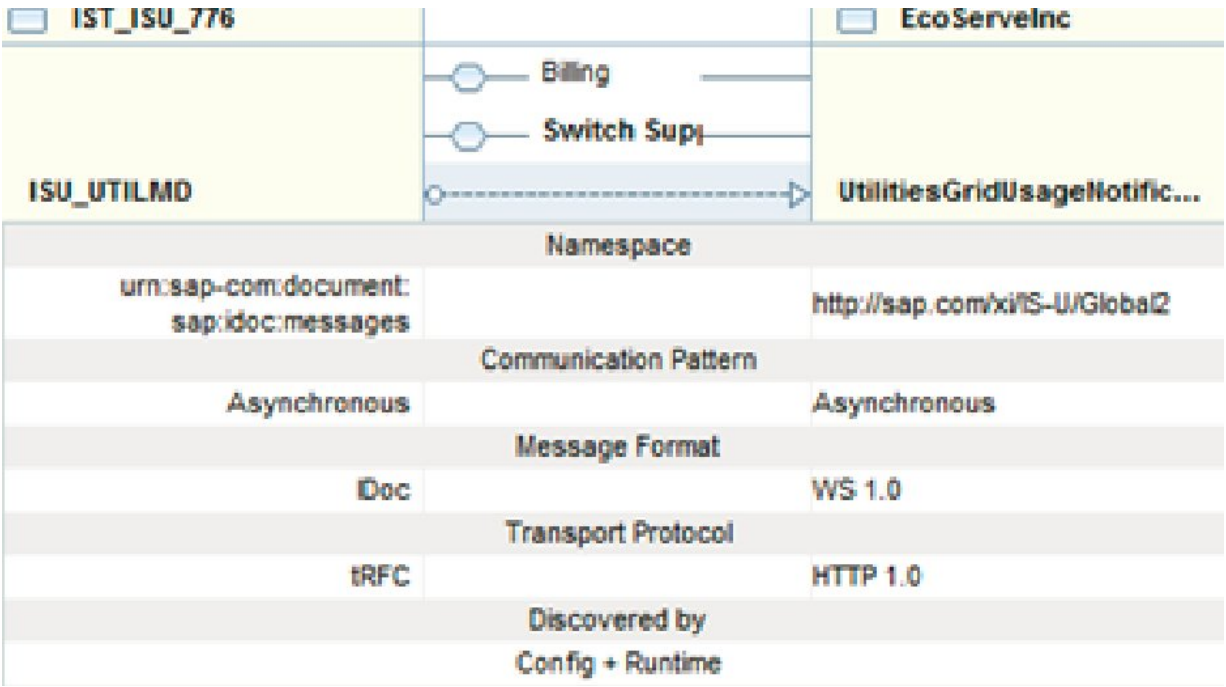

Fig. 3. Conversation details of a message flow

For instance, the following queries specify a keyword search with search term and restriction of the type in the result set to *Host* (i.e., the physical machine on which business applications are running):

```
http://localhost/search
    ?query=term\&type=Host\&...
```

and *field*, as field specific search criteria:

```
http://localhost/search
    ?location=Sydney.
```

In the same way, the result set can be defined to return any information in the BN by traversing the network e.g. from a specific participant *system1* (i.e., an application or tenant),

```
http://localhost/SYSTEM1/
    ?show=meta,location,host.name,
```

which returns *location* information of the participant itself and the *name* of the connected host the participant runs on. Simple Friend of a Friend (FoaF) queries returning, e.g., the hosts of all neighbors of the participant equally straight forward:

```
    http://localhost/SYSTEM1/neighbors/
  host/.
```

Due to the decoupling of the data query and traversal components from the network inference and through model-centric index generation, all requests are processed within short time even on larger networks (see [14] for performance numbers).

### *B. Update Processing in BNS*

The BNS distinguishes three types of updates, (a) the steady loading of raw data from known and new data sources, which affect the already computed BN from the inference direction (see *REQ-5*), (b) the systematic enrichment (i.e., labeling, grouping w/o effect on the data source) and (c) enhancement (i.e., re-deployment, possibly affects the data sources) through the GUI and client APIs (see *REQ-4*).

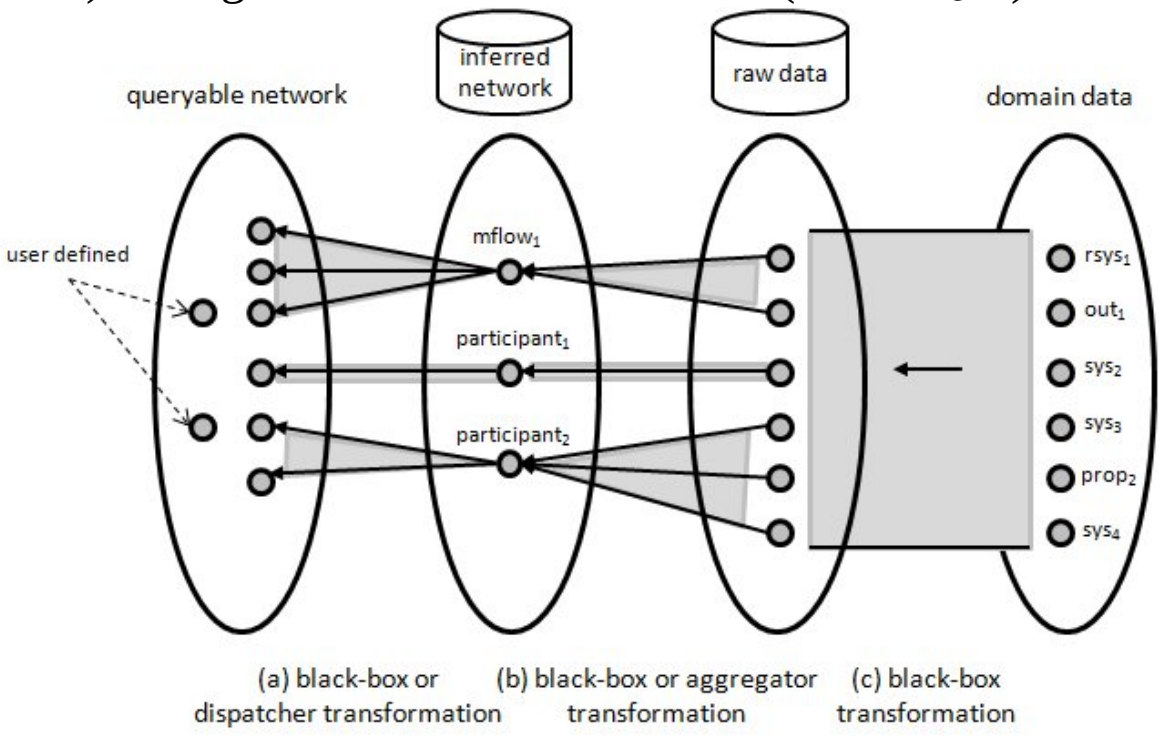

Fig. 4. Data Provenance from domain data to the queryable network with transformation classes

For case (a), Figure 4 shows the described BNS architecture with the data streams and their transformations (categorized according to [7]). The data from the sources (e.g., $rsys_1$, $out_1$, ..., $prop_2$, $sys_4$) is pushed without transformation (black-box) to the inference system, which stores the conform raw data, while keeping potential duplicate information. However, the unique identifiers from the source systems may not be unique in the inference system. To avoid "collisions" in case identifier occur more than once across different sources, the records are stored with a composed key containing their locally unique identifier and their origin. Keeping this in mind, records from the same origin with the same identifier are updated (i.e., remain as one record), while all other cases lead to different records (i.e., same origin, different keys; same key, different origin). That means, if a record *sys("h7", "myH7", "originH7")* is pushed to the inference system, which already contains a record *sys("h7", "", "originH7")* with the same key and origin, the records are already merged in the inbound to *sys:("h7", "myH7", "originH7")*. In case of records without any primary key, the identity of the information cannot be determined. Hence a hash function is calculated over the record values, which leads to new records whenever the information is loaded. It then is the task of the inference programs to identify equivalence between the records and chose a meaningful surrogate. There are cases like for *same_sys("h7", "h8")* (i.e., an equivalence relation between logical applications or tenants), in which the record has more than one identifiers. These are simply represented as combined primary key. The lineage tracing for the black-box transformation leverages the unique keys within the source combined with the origin information, which directs to the correct source. In this way, an anchor to the sources is created, which however lies in the storage close to the inference system (see *REQ-4*, *REQ-5*).

One of the tasks of the inference programs is to find equivalences between the same entities of the BN model. Due to the nature of equivalence classes, the most common transformations to the BN (i.e., inferred network) are either black-box or aggregators (see Figure 4). The black-box transformations translate one record in the raw data to one record in the BN, which allows for direct lineage tracing (see *REQ-4*). Since the identifiers in the BN become immediately visible to the applications, new keys are generated and a key mapping is maintained (e.g., from the record *sys("h7", "myH7", "originH7")* to the BN record *participant("1", "myH7")* the mapping information is as follows <*sys:h7, participant:1*>).

The more difficult case is the aggregation of information. For instance, Figure 4 shows the reconstruction of the message flow $mflow_1$ from $rsys_1$ and $out_1$ raw data, the black-box transformation from $sys_2$ to $participant_1$ (as discussed before) and the aggregation from $sys_3$ and $sys_4$ extended by a fragment $prop_1$, which might carry complementing information, to $participant_2$. For the latter, at least three variants are possible: (1) perform a "destroying merge" that identifies a leading set of data (the leading object or surrogate) and enrich the record by missing information from equivalent objects (e.g., add description "myH7" to the leading object *sys("h7", "", "originH7")*) and update the surrogate by any change from the sources, (2) perform a "preserving merge", which keeps all equivalent records and identifies a surrogate filled up with missing information (similar to (1)), while remembering from which object the information in the surrogate came from, or (3) do not aggregate at all, but make the equivalence information public for the applications and let them handle the merge. Option (3) is clearly the simplest one for the inference system, since it provides all information to the caller, but leaves it with the task of handling equivalences. The major drawback for the BNS is the computation of one connected network, which becomes difficult if equivalences cannot be addressed by only one surrogate. The most extreme alternative to that is (1), which makes the definition of connected components easier, while making update processing from the sources and lineage tracing from the BN (nearly) impossible. The "information preserving", surrogate approach (2) comes with the highest data management efforts, but fulfills *REQ-4* to the BNS best. At all times, the lineage tracing down to the sources (i.e., for operations on discovered records) and the steady integration of new sources and updates is granted. For instance, if a new source is added, which adds a system with key $sys_6$ equivalent to $sys_5$, then it is simply added to the equivalence class and the surrogate is re-calculated based on the updated information. However, option (2) has some major disadvantages besides its complexity, upon them (i) finding a function that calculates a good surrogate, (ii) the constant re-evaluation of the surrogate, and (iii) removing records that contributed nearly exclusively to the surrogate. In the current BNS all of them are mitigated intermediately, but require further analysis and leave room for further research. To sketch some ideas, approach (i) currently takes the most complete record from the equivalence class and copies its values to the surrogate. If this record was deleted from the source (iii), then the second record is chosen and so on, until the equivalence class is empty and the object is removed from the network. The major issue with (iii) is the user's experience, when between two loads, some objects in the BN provide less information or maybe cannot be found any more due to sudden lack of information. Currently this can only be prevented by good user enrichment (e.g., labeling). The steady re-evaluation (ii) cannot be avoided, but only optimized (e.g., by a delta-calculation technique, which only re-evaluates the fields that have changed).

The enrichment of BNs (b) stands for non-modifying operations to the network data like adding labels to network instances or grouping them for more intuitive visualization (see *REQ-3*). These operations lead to artifacts stored along with the network, which are attached to the leading object (e.g., of a participant equivalence class), which determines the artifact's lifecycle. Figure 4 shows them in the queryable network space, which means that they are treated in the same way as the computed instances (i.e., indexed, searchable and traversable), however they have their origin on the surface and won't be deployed into the known data sources.

The enhancement of the BN (c) is treated differently in terms of re-deployment to the sources (see *REQ-3*). For that, computed BN instances (e.g., participants, message flows) are modified, newly added or deleted. These changes can then be pushed back to the respective source systems. Clearly, the operations on existing instances (e.g., modify, delete) require a bijective mapping from the source to the raw data, and a stringent provenance from the inference model down to the sources. For the surrogates this requires a good book keeping on how the leading object's information was selected. For instance, if more than one of the equivalent records has a non-empty description, which they could contribute to the surrogate, the evaluation function has chosen only one of the two. In case of changes to the description, the origin can be clearly identified and re-deployed down to source system. With this approach, there are several issues, which are left open for further research. The BNS treats them according to the following leading principles:

- Enhancements from the user are valued higher than the information from the automatic discovery and replace the discovered information within the surrogate (refer to [15] for the theory of coming from the computed "as-is" to a "to-be" network through expert knowledge, i.e., enhancements)
- The enhancements can be re-deployed to the sources if sufficient lineage tracing exists (i.e., for the creation of new instances, e.g., a participant, the source has to be specified and enough domain specific information has to be provided to create and activate the instance in the source system. These sources can be middleware system, for which the new instance should be runnable as any other created locally. Since the inference system does not know about source domain specificities, the entered information is passed through to the sources without any further checking)
- The function that calculates the surrogate is supported by a configuration, which allows the user to order the type

of source (e.g., system landscape, runtime, configuration) by relevance (i.e., more grain granular) or even on instance level, in which order the information from the equivalent records are considered.

The update processing from the client API works again from simple `HTTP POST`, `PUT`, `DELETE` or through the native Java/ OSGi API.

## IV. Related Work

For the overall system approach, related work is conducted in the area of Process Mining (PM) [1] and [2], which sits between computational intelligence and data mining. It has similar requirements for data discovery, conformance and enhancement with respect to NM [13], but does not work with network models and inference. PM exclusively strives to derive BPM models from process logs. Hence PM complements BNM in the area of business process discovery.

Gaining insight into the network of physical and virtual nodes within enterprises is only addressed by the *Host* entity in NIM, since it is not primarily relevant for visualizing and operating integration networks. This domain is mainly addressed by the IT service management [11] and virtualization community [6], which could be considered when introducing physical entities to our meta-model.

The linked (web) data research, shares similar approaches and methodologies, which have so far neglected linked data within enterprises and mainly focused on RDF-based approaches [4] and [5].

A distantly related field is the work on software ecosystems [10], in which the notion of interconnected enterprises is defined similar to BN, but reduced to the common software development process. Hence the approaches for modeling [9] and governance [3] of these ecosystems do rather complement, but do not overlap.

## V. Discussions And Future Work

In this work, we present insights into a reference implementation of the query and data processing within a Business Network System based on the theory on Business Networks [13] and Business Network Management [15] as well as the reference architecture discussed in [16]. For that, we added a client API [14] and data provenance approach to complement the BNS towards an emergent, enterprise-ready system. The architecture constitutes a holistic network data management platform, reaching from the information retrieval, network mining and inference, up to the query and traversal on the network data, while providing some insights and solutions for the continuous source data load as well as user enhancements through data provenance.

The data provenance topic for the re-deployment of user information to the sources requires further research. Secondly, opportunities for performance improvements lie in the selection of the databases. Significant performance gains are expected by running the BNS on the SAP HANA computing engine. Early prototypes show promising results.

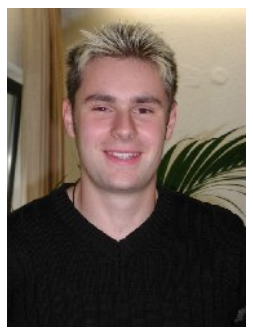

**Daniel Ritter** received a M.Sc in computer science and mathematics in 2008 from the University of Heidelberg, Germany. He is currently working as Research and (Software) Development Architect (VP level) in Technology Development with the department of Process- and Network Integration at the SAP AG, Walldorf, Germany. His current research interests include network mining and reconstruction, logic programming, computer language design and compilation, databases, and data management.